\documentstyle[aas2pp4,psfig]{article}

\lefthead{Kuulkers, van der Klis, \&\ Parmar}
\righthead{4U\,1630$-$47}

\begin{document}

\title{Evidence for an anomalous state in the black-hole candidate 4U\,1630$-$47}

\author{Erik Kuulkers\footnote{Present address:
	Astrophysics, Nuclear and Astrophysics Laboratory, Keble Road,
	Oxford OX1 3RH, United Kingdom; Electronic mail: 
	e.kuulkers1@physics.oxford.ac.uk}}
\affil{ESA/ESTEC, Astrophysics Division, P.O.~Box 299, 2200~AG
	Noordwijk, The Netherlands; \\
	Electronic mail: ekuulkers@astro.estec.esa.nl}

\author{Michiel van der Klis}
\affil{Astronomical Institute ``Anton Pannekoek'', University of Amsterdam
and Center for High Energy Astrophysics,
Kruislaan 403, 1098 SJ Amsterdam, The Netherlands\\
Electronic mail: michiel@astro.uva.nl}

\and

\author{Arvind N. Parmar}
\affil{ESA/ESTEC, Astrophysics Division, P.O.~Box 299, 2200~AG
	Noordwijk, The Netherlands; \\
	Electronic mail: aparmar@astro.estec.esa.nl}

\begin{abstract}

We have re-analyzed two EXOSAT Medium Energy observations during part of
the 1984 X-ray outburst of the black-hole candidate transient system 
4U\,1630$-$47.
One observation (May 10) shows fast timing properties 
(on timescales shorter than $\sim$250\,s) typical of those of other black 
hole candidates in their high states.
The power-spectrum shows a power law with index $\sim$0.7 and fractional rms 
amplitude (0.001--1\,Hz) of $\sim$2\%.
Another observation, obtained one month earlier (April 11), shows different 
fast timing behavior. The power spectrum of this observation can be 
described by a power law with index $\sim$1.5 and fractional rms amplitude
(0.001--1\,Hz) of $\sim$7\%. 
This is inconsistent with the behavior during low, high or very high states 
known to occur in most of the black hole candidates.
Instead, the timing properties resemble those recently reported 
from ASCA observations of the superluminal black-hole candidates 
GRS\,1915+105 and GRO\,J1655$-$40 (X-ray Nova Sco 1994). 
This timing behavior may represent a new state of black hole candidates.

In addition, the X-ray spectra of the three sources show similar properties.
Since GRS\,1915+105 and GRO\,J1655$-$40 exhibit superluminal jets, we
propose that 4U\,1630$-$47 is also a relativistic jet source.

\end{abstract}

\keywords{accretion, accretion disks --- binaries: close ---
stars: individual (4U\,1630$-$47) --- black hole physics --- X-rays: stars}

\section{Introduction}

4U\,1630$-$47 is an ultra-soft X-ray transient, which has shown multiple 
outbursts. It was discovered by Jones, Forman, \&\ Tananbaum  (\cite{jft76}), 
and has the shortest known outburst recurrence interval of $\sim$600~days 
(e.g., Parmar, Angelini, \&\ White \cite{paw95}; Parmar et al.\ \cite{pwk96}). 
In general, the outbursts rise within a few weeks and decay on an e-folding 
time scale of $\sim$50~days. The outburst duration is variable with some 
outbursts having durations of up to $\sim$2.4 years 
(Kaluzienski \&\ Holt \cite{kh77}; Share et al.\ \cite{swb78}; 
Kaluzienski et al.\ \cite{kbh78}; Sims \&\ Watson \cite{sw78}; 
Kitamoto \cite{k96}).

Based on its observed properties Parmar, Stella, \&\ White (\cite{psw86})
proposed that 4U\,1630$-$47 should be regarded as a black-hole candidate. The 
X-ray spectrum has two components, an (ultra) soft component and a (ultra)
hard power-law tail. As an outburst proceeds, the soft component fades 
and the hard component becomes relatively more intense. Close to outburst 
maximum, strong variability is observed with a characteristic time scale
of $\sim$20\,s. No pulsations or bursts have been detected.
These properties are similar to those of other black-hole candidates
(see, e.g., Tanaka \&\ Lewin \cite{tl95}).

Recent studies have revealed that the black-hole candidates show
three different states, mainly based on the characteristics in the 
X-ray spectral and power spectral behavior (see Tanaka \&\ Lewin \cite{tl95}, 
Van der Klis \cite{k95}, and references therein).
These are the so-called low, high and very high states.
The three states are thought to be governed by the mass accretion
rate. During the low state, the power spectra are characterized by
a power-law with index $\alpha$$\sim$1--1.7. Below $\sim$0.1\,Hz
the power spectrum flattens and above $\sim$10\,Hz it steepens. 
This shape is called band-limited noise.
The fractional rms amplitude of the noise is large, typically between
30 and 50\%.
Quasi-periodic oscillations (QPOs) in this state have been reported 
with centroid frequencies in the range 0.04--0.8\,Hz. During this
state the X-ray spectrum is dominated by the hard power-law component.
The high state is characterized by weak power-law noise with $\alpha$$\sim$1
and fractional rms variability of a few percent.
QPO at $\sim$0.08\,Hz with a harmonic, and fractional rms up to $\sim$3\%, 
were seen in LMC\,X-1 (Ebisawa et al.\ \cite{emi89}). 
During this state the X-ray spectrum is dominated by the soft component.
In the very high state the power spectra show 3--10\,Hz QPO and
rapidly variable broad-band noise.
The X-ray spectrum is similar to the high state spectrum scaled
to higher intensities, although the hard power law is slightly steeper
(index $\sim$2.5 in the very high state versus $\sim$2 in the
high state, see e.g., Ebisawa et al.\ 1994).

Interest in black-hole transients has been heightened by the 
discovery of superluminal jets in the two Galactic transient sources
GRS\,1915+105 and GRO\,J1655$-$40/X-ray Nova Sco 1994 
(see, e.g., Mirabel \&\ Rodr\'{\i}guez \cite{mr96}, for a recent review). 
It was argued by Mirabel \&\ Rodr\'{\i}guez (\cite{mr94}) after their 
discovery of the jets in GRS\,1915+105, that other black-hole candidate 
transients might also exhibit relativistic jets (see also Hjellming \&\ Rupen 
\cite{hr95}). Bailyn et al.\ (\cite{bog95}) presented dynamical 
evidence that the compact object in GRO\,J1655$-$40 has a mass
appropriate to a black hole.

Recently, Ebisawa (\cite{e96}) presented the timing behavior of GRO\,J1655$-$40
as observed on two occasions (see also Zhang et al.\ \cite{zes96}) and of
GRS\,1915+105 as observed on one occasion, using ASCA. The
power spectra were characterized by a featureless power law
with power-spectral indices between 1.2--1.7 and fractional rms 
amplitudes of 6--8\%\ in the 0.001 to 1\,Hz frequency range. 
As we will show, such a shape does not fit in with the typical black-hole 
candidate behavior described above.

We report on EXOSAT observations made on two occasions 
during the 1984 outburst of 4U\,1630$-$47. This outburst was discovered by 
{\it Tenma} (Tanaka et al.\ \cite{txx84}). Energy spectra and 
auto-correlation functions 
of the EXOSAT data were reported in Parmar et al.\ (\cite{psw86}). 
We demonstrate that during one of the observations, 4U\,1630$-$47 exhibited
temporal variability which is different from the variability in the 
three black hole candidate states described above.
We discuss this behavior in the light of the recent ASCA timing 
results of the two newly discovered X-ray transients GRS\,1915+105 and 
GRO\,J1655$-$40. Our analysis provides evidence that 4U\,1630$-$47 belongs 
to the class of black-hole candidates. 

\section{Observations and analysis}

We analyzed the EXOSAT Medium Energy (ME) experiment argon data
(Turner, Smith, \&\ Zimmerman \cite{tsz81}; White \&\ Peacock \cite{wp88}) 
of two observations made on 1984 April 11 and May 10
(see also Parmar et al.\ \cite{psw86}). 
In Table~1 we present an observation log, together
with the instrument set-up. For details concerning the 
data modes we refer to, e.g., White \&\ Peacock (1988).
During the two further EXOSAT observations reported in Parmar et al.\ 
(\cite{psw86}), the source was too faint to perform a significant 
temporal analysis.

The fast timing behavior of 4U\,1630$-$47 was studied by 
performing fast Fourier transforms (FFTs) on successive 256\,s blocks of 
the HTR3 data. This resulted in 81 and 54 FFTs for the April and May 
observations, respectively.
The white noise level, or Poisson level, 
was subtracted from the data. We estimated the Poisson level, modified
by instrument dead-time, using the results obtained from the study of
dead-time effects on HTR data (Berger \&\ van der Klis \cite{bk94}). 
The study of the HTR data also revealed the presence of an instrumental
band-limited noise component, with a cut-off frequency of $\sim$100\,Hz and 
a fractional rms amplitude of $\sim$3\%\ of the total observed count rate
(Berger \&\ van der Klis \cite{bk94}, \cite{bk96}). This component was also 
subtracted from the power spectra. 

In this paper we describe the X-ray power density spectrum as a sum of
two noise components, a power-law shaped very-low-frequency noise (VLFN)
($P_{\rm VLFN}(\nu) = A_{\rm V}\nu^{-\alpha_{\rm V}}$,
where $\nu$ is the frequency, $\alpha_{\rm V}$ the power-law index, and
$A_{\rm V}$ the normalization constant), and a Lorentzian-shaped QPO
($P_{\rm QPO}(\nu) = A_{\rm Q} [(\nu - \nu_{\rm Q})^2 + (\Delta\nu_{\rm Q}/2)^2]^{-1}$,
where $\Delta\nu_{\rm Q}$ is the full width at half maximum (FWHM) of the
QPO and $A_{\rm Q}$ a normalization constant).
The strengths of these noise components are expressed in terms of the
fractional rms amplitudes of the corresponding fluctuations in the time
series. They are calculated by integrating their contributions to the power
spectra over certain frequency ranges, as determined from fits of the
functional shapes to the power spectra. The VLFN integration range was 
0.001 to 1\,Hz. All errors on the quoted power spectral parameters
were determined from an error scan through the $\chi^2$ space using 
$\Delta\chi^2=1$.

\section{Results}

The average count rates (1--20\,keV) during the April and May 
observations of 4U\,1630$-$47 were 
$\sim$0.8 cts\,s$^{-1}$\,cm$^{-2}$ and 
$\sim$0.3\,cts\,s$^{-1}$\,cm$^{-2}$, corresponding to
$\sim$1$\cdot$10$^{-8}$\,erg\,s$^{-1}$\,cm$^{-2}$ and 
$\sim$4$\cdot$10$^{-9}$\,erg\,s$^{-1}$\,cm$^{-2}$, respectively.
During the April observations the intensity was extremely variable,
while during the May observation there was almost no intensity variability 
(see Fig.~3 of Parmar et al.\ \cite{psw86}).

The average power spectra of these two observations are shown in Fig.~1a. 
It is clear that the two power spectra are different. We fit the power 
spectra to the functional forms described in Sect.~2. These fit results 
are shown in Figs.~1b,c. The April power spectrum showed strong
VLFN up to $\sim$5\,Hz, with a fractional rms of 6.7$\pm$0.2\%\
and a power-law index of 1.50$\pm$0.02 ($\chi^2$=69.1 for
35 degrees of freedom).
The May power spectrum could be represented by a power-law
with an index of 0.65$\pm$0.05 and a fractional rms amplitude of 
1.8$\pm$0.1\%\ ($\chi^2$=32.8 for 35 degrees of freedom). 

There is evidence for excess variability around $\sim$1\,Hz in the May 
power spectrum. Although a power law described the data satisfactorily
(see above), we also included a QPO in our fits (Fig.~1d). 
We find that if there is QPO present, it has a fractional rms of
1.6$\pm$0.4\%, a FWHM of 0.6$\pm$0.5\,Hz, and
centroid frequency of 1.02$\pm$0.13\,Hz
($\chi^2$=20.2 for 32 degrees of freedom).
In this case, the VLFN has a power-law index of 0.74$\pm$0.12
and fractional rms of 1.7$\pm$0.1\%. From an F-test for 
the additional QPO term and a 90\%\ confidence error-scan
of the integrated power in $\chi^2$-space ($\Delta\chi^2$=1), 
we conclude that this QPO is significant 
at a level of $\sim$2.5$\sigma$.

\section{Discussion}

We have demonstrated that on two occasions during the 1984 outburst, 
4U\,1630$-$47 showed different fast timing behavior.

During the first observation in April, near the peak of the outburst,
the intensity showed a strong variability down to seconds, whereas during
the May observations the variability was less strong, as already 
pointed out by Parmar et al.\ (\cite{psw86}).  
We investigated these intensity changes by computing FFTs of the 
data. The power spectra during the April observation showed
a VLFN power law with index $\alpha$$\sim$1.5 and fractional rms amplitude 
of $\sim$7\%. The strength is comparable with the rms excess variability 
in the 1--7\,keV band reported by Parmar et al.\ (\cite{psw86}). 
The power spectra of the May observation, however, showed a VLFN power law 
component with a fractional rms of only $\sim$2\%, and a flatter slope, 
$\alpha$$\sim$0.7. Although, formally the May power spectrum can be 
satisfactorily fit by a power law, we find evidence 
(at a $\sim$2.5$\sigma$ level) for the presence of QPO
at $\sim$1\,Hz, with fractional rms of $\sim$1.6\%\ and FWHM of $\sim$0.6\,Hz.
During both the April and May observations, the soft spectral component
dominated the X-ray spectra (Parmar et al.\ 1986).

The timing properties of the May observation are reminiscent of those
of other black hole candidates in their high states, where power law VLFN
with indices $\alpha$$\sim$1 and small fractional rms amplitudes (few \%\/)
are commonly observed. In this state the X-ray spectra are normally 
dominated by the soft component in the 2--10\,keV band. We note that QPO 
in the high state have also been reported by Ebisawa et al.\ (\cite{emi89}) 
in LMC\,X-1 at a frequency of $\sim$0.08\,Hz.

Parmar et al.\ (\cite{psw86}) classified the spectral evolution 
of 4U\,1630$-$47 from 1984 April to July as consistent with 
a transition from high to low state. This suggests,
that the source was also in the high state during
the April observation. However, as we have shown above, the April power spectra
are different from the May (high state) power spectra and different from 
power spectra generally seen for black-hole candidates in their high 
state, and we therefore conclude that during the April observations 
4U\,1630$-$47 was not in a high state\footnote{It was already noted by 
Parmar et al.\ 1986, that the rapid variability during the April observations 
was stronger than that seen during the high state of Cyg\,X-1.}.
Black-hole candidates in their low state show very 
strong band-limited noise (up to $\sim$50\%) and hard power law X-ray spectra.
Since the April power spectrum shows only moderately strong 
($\sim$7\%\/) power-law noise, together with an (ultra) soft X-ray spectrum, 
the source was also unlikely to be in a low state.

Up to now, only two sources have been seen in the very high state, 
i.e., GX\,339$-$4 and GS\,1124$-$68 (Miyamoto et al.\ \cite{mik93},
Van der Klis 1995, and references therein). 
In the very high state the power spectra show rapidly variable broad-band 
noise with amplitudes between 1 and 15\%, flat tops below 
0.05--10\,Hz and power-law shapes with index $\sim$1 above these
frequencies. The broad-band noise becomes stronger at higher energies.
Most of the time 3--10\,Hz QPO are present during the 
very high state. The April power spectrum differs from these very high state 
variability characteristics:
it has a featureless steeper ($\alpha$$\sim$1.5) power law shape, with no 
indication for a flat top. Moreover, Parmar et al.\ (1986) found that 
the variability at
high energies (14--30\,keV) is weaker than that at lower energies
(1--7\,keV), i.e., $\sim$4\%\ versus 7\%, respectively, which is different from
the very high state broad-band noise energy dependence.
So, we conclude that the observed timing behavior does not fit 
in with the characteristic timing behavior of either the low state, the 
high state, or the very high state.

GRS\,1915+105 (Castro-Tirado et al.\ \cite{cbl94}) and GRO\,J1655$-$40 
(Harmon et al.\ \cite{hwz95}) are recently discovered X-ray transients,
of which the latter has dynamically been established to contain a black hole
(Bailyn et al.\ 1995).
Ebisawa (\cite{e96}) presented the power spectral 
behavior of these two X-ray transients from data observed with ASCA. 
The power-spectral shapes during the observations of GRO\,J1655$-$40
(see also Zhang et al.\ \cite{zes96})
on two occasions (end 1994 September and mid 1995 October) and 
of GRS\,1915+105 (end 1994 September) were consistent with a single power law
with indices 1.2--1.7 and fractional rms amplitudes of 6--8\%\ in the
0.001--1\,Hz frequency range.
Moreover, during the recent (soft) X-ray outburst of GRO\,J1655$-$40
(Remillard et al.\ \cite{rbc96}) similar strong variations on timescales of 
2\,s--10\,min (0.002--0.5\,Hz) were found (5--10\%\/) in data obtained with
RXTE (Horne et al.\ \cite{hhb96}).
These ASCA power spectra are very similar in shape and 
strength to the power spectra of the April EXOSAT observation
of 4U\,1630$-$47. Since they are all inconsistent with the 
low, high and very high state behavior usually observed (see above), 
we suggest that this behavior represents an anomalous state in black-hole 
candidates. 
We note, however, that recent RXTE observations of GRS\,1915+105 
(Greiner, Morgan, \&\ Remillard \cite{gmr96}) revealed that this source 
also shows several other power spectral shapes. 
Close monitoring and a detailed intercomparison of the outbursts of
GRS\,1915+105, GRO\,J1655$-$40 and 4U\,1630$-$47 with, e.g., RXTE are 
therefore needed in order to make sure that the anomalous timing behavior 
represents a completely new state in (black-hole) X-ray binaries.

During the April observations 4U\,1630$-$47 exhibited an X-ray spectrum 
which had an ultrasoft component together
with a hard power law with index $\sim$2.5 (Parmar et al.\ 1986).
Both GRS\,1915+105 and GRO\,J1655$-$40 were bright during the 1994 September 
ASCA soft X-ray observations, with indications of spectral softening at 
high energies (Nagase et al.\ \cite{nik94}), while the BATSE hard X-ray 
(20--100\,keV) flux was at near quiescent values (Harmon et al.\ \cite{hwz95}; 
Paciesas et al.\ \cite{pdh95}). Similar spectral behavior
was reported a few days before these ASCA observations from TTM and 
HEXE data by Alexandrovich et al.\ (\cite{abe94}).
During the August 1995 ASCA observations GRO\,J1655$-$40 was a factor
$\sim$3 brighter in the ASCA band (Inoue, Nagase, \&\ Ueda \cite{inu95}), 
and the BATSE observations showed a steep hard X-ray spectrum with a 
spectral index near 2.5 (Zhang et al.\ \cite{zhp95}).
This shows that, apart from the power spectral behavior, the X-ray spectral 
behavior was also similar in the three sources. 

We note that these X-ray spectral properties
are consistent with the general X-ray spectral behavior in the 
very high state in black-hole candidates (see Miyamoto et al.\ \cite{mkk91};
\cite{mik93}, \cite{mki94}; Ebisawa et al.\ \cite{eoa94}):
the X-ray intensity in the very high state is 2--3 times higher compared to 
the high state, and the X-ray spectra are dominated by the (ultra) soft 
component; the high energy part of the X-ray spectra ($\gtrsim$10\,keV) is
described by a power law with index $\alpha$$\sim$2.5, i.e., slightly steeper 
than the high energy spectrum in the high state ($\alpha$$\sim$2).
As suggested by Zhang et al.\ (1996), this may imply that the observed 
anomalous timing properties are not related to X-ray spectral components, but 
to the nature of these sources. 
Since GRS\,1915+105 and 4U\,1630$-$47 have not
(yet) been dynamically shown to contain a black hole, we do not rule
out the possibility, that the observed 
timing behavior may be unconnected to whether the compact object
is a neutron star or a black hole.

Since the fast ($\lesssim$250\,s) timing properties of 4U\,1630$-$47 
during the April observations (power-law shaped power spectra with index 
$\sim$1.5 and fractional rms of $\sim$7\%\/) and the X-ray spectra 
(ultra soft component and hard power-law component with index $\sim$2.5) 
are very similar to that 
observed from the sources GRS\,1915+105 and GRO\,J1655$-$40, we propose that 
these sources have something in common with 4U\,1630$-$47,
which is not present in other black-hole candidates.

Episodic ejections of relativistic, apparently superluminal, jets have been
observed from GRS\,1915+105 and GRO\,J1655$-$40 during radio outbursts 
(Mirabel \&\ Rodr\'{\i}guez \cite{mr94}; Tingay et al.\ \cite{tjp95};
Hjellming \&\ Rupen \cite{hr95}). Observations in hard X-rays with BATSE 
(Hjellming \&\ Rupen \cite{hr95}; Harmon et al.\ \cite{hwz95};
Paciesas et al.\ \cite{pdh95}) suggest that the radio ejections
may occur in conjunction with the hard X-ray outbursts
(see also Meier \cite{me96}), although no simple one-to-one relationship 
does exist (see e.g.\ Zhang et al.\ 1996).
We note that during the 1994 September ASCA observation of GRO\,J1655-40
the radio flux was declining after a second radio peak with ejections
$\sim$10 days earlier (see, e.g., Hjellming \&\ Rupen \cite{hr95}),
and that recently the radio flux from GRO\,J1655-40 has increased again 
(Hunstead \&\ Campbell-Wilson \cite{hc96})
during the current (soft) X-ray outburst (see above).

If our proposal that 4U\,1630$-$47 is a similar to the superluminal
sources GRS\,1915+105 and GRO\,J1655$-$40 is true, we speculate that during 
the 1984 outburst of 4U\,1630$-$47 relativistic radio jets might have been
present during or near the outburst.
It is interesting to note that changes in the low-energy absorption
observed by Parmar et al.\ (\cite{psw86}) as the 1984 outburst
evolved could indicate the presence of outflowing material 
(Levinson \&\ Blandford \cite{lb96}).
Whether relativistic radio jets are present or not during other outbursts of 
4U\,1630$-$47 can be tested during its current outburst 
(Marshall \cite{ma96}; Levine et al.\ \cite{lbc96}) or at the time of the 
next outburst, which is expected to occur around 1997 October using the 
ephemeris of Parmar et al.\ (\cite{pwk96}).

\acknowledgments

EK acknowledges receipt of an ESA Fellowship, and MvdK support from
ASTRON. We thank two anonymous referees for their valuable 
comments which improved this Letter.

~\\
~\\
{\bf Note added in proof:} We note that the 1996 outburst of 4U\,1630$-$47, 
which started in March (Marshall 1996; Levine et al. 1996), ended near 
the beginning of August, when the intensity dropped below the RXTE ASM 
detection limits.

\newpage

\begin{table}[h]
\begin{center}
\begin{tabular}{clcccccc}
\multicolumn{8}{c}{\normalsize{\bf Table 1:} EXOSAT ME Argon observation log of 4U\,1630$-$47} \\
\multicolumn{8}{c}{~} \\
\hline\hline
\multicolumn{1}{c}{Year} & \multicolumn{1}{c}{Date} &
\multicolumn{1}{c}{Day$^a$} & \multicolumn{1}{c}{UT start} &
\multicolumn{1}{c}{UT end} & \multicolumn{1}{c}{Mode$^b$} &
\multicolumn{1}{c}{Time} & \multicolumn{1}{c}{Config.$^c$} \\
\multicolumn{1}{c}{~} & \multicolumn{1}{c}{~} &
\multicolumn{1}{c}{~} & \multicolumn{1}{c}{(hr:min)} &
\multicolumn{1}{c}{(hr:min)} & \multicolumn{1}{c}{~} &
\multicolumn{1}{c}{Resolution} & \multicolumn{1}{c}{~} \\
\hline
1984 & Apr 11 & 102 & 08:08 & 15:10 & E4/E5/T3$^d$ & 10\,s/0.1\,s/8\,ms & WA \\
1984 & May 10 & 131 & 02:00 & 06:45 & E5/E6/T3$^e$ & 0.6\,s/20\,ms/8\,ms & WA \\
\hline
\multicolumn{8}{l}{\scriptsize $^{a}$Jan 1 = day 1.} \\
\multicolumn{8}{l}{\scriptsize $^{b}$E = HER mode = High Energy Resolution mode, T = HTR mode = High Time Resolution mode.} \\
\multicolumn{8}{l}{\scriptsize $^{c}$Array configuration on source: WA=whole array.} \\
\multicolumn{8}{l}{\scriptsize $^{d}$From 08:08 -- 13:16 HER4/HTR3, from 13:39 -- 15:10 HER5/HTR3.} \\
\multicolumn{8}{l}{\scriptsize $^{e}$From 02:00 -- 02:32 HER5/HER6, from 02:38 -- 06:45 HER5/HTR3.} \\
\end{tabular}
\end{center}
\end{table}


\centerline{\hbox{
\psfig{figure=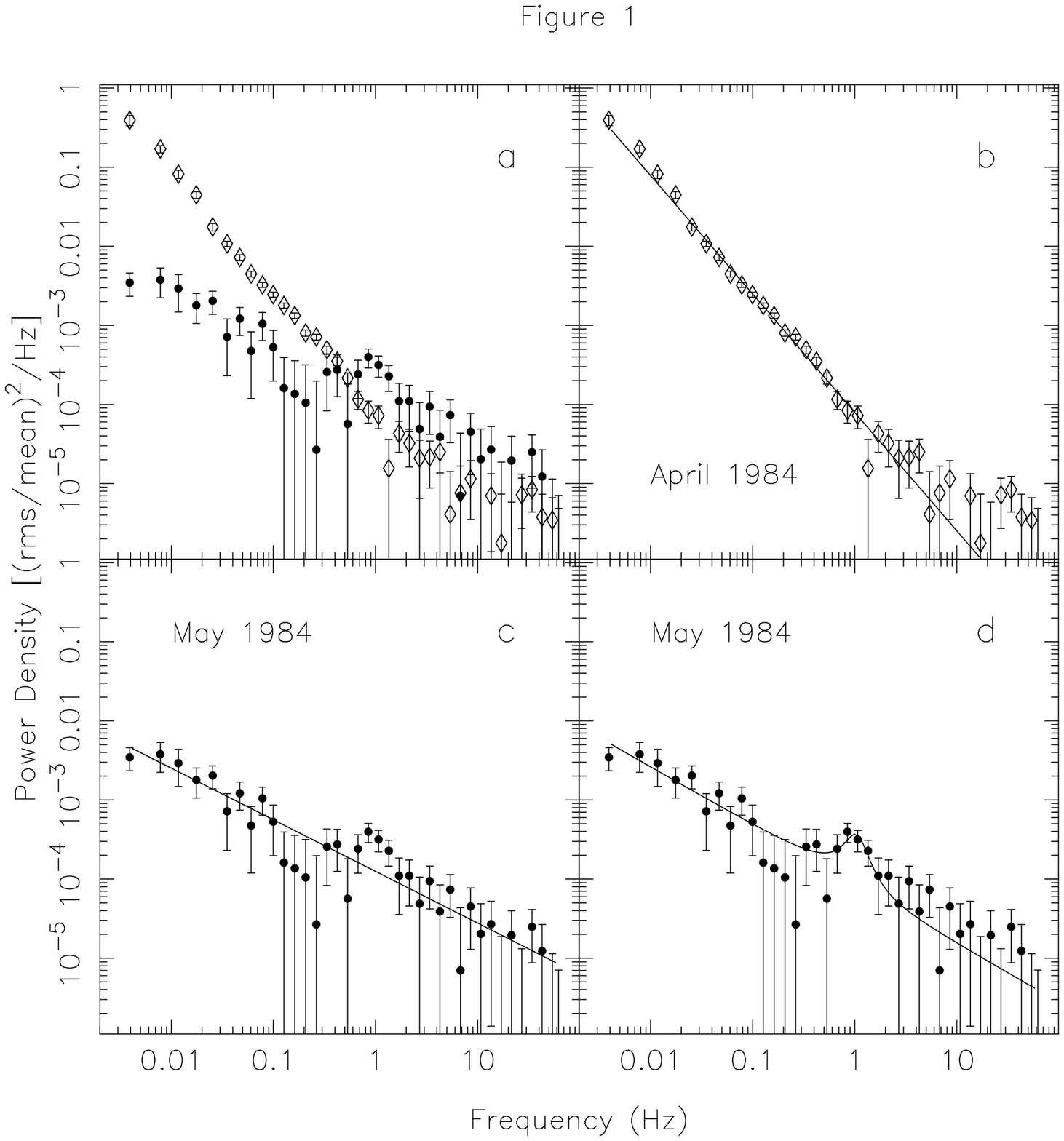,height=12.2cm,bbllx=34pt,bblly=109pt,bburx=571pt,bbury=610pt,angle=-90}
}}

\noindent
{\bf Figure 1.} ({\em a}) Average power spectra for the 1984 April ($\Diamond$)
and 1984 May ($\bullet$) observations. ({\em b}) The 1984 April power
spectrum with the best fit power-law component. ({\em c}) The May 1984
power spectrum with the best fit power-law component. ({\em d})
Same as ({\em c}) with the inclusion of the QPO component.

\end{document}